\documentclass[11pt]{elsarticle}
 
\usepackage{calrsfs,graphicx,array,natbib,enumitem}
\usepackage{setspace,wrapfig,marvosym,bbm,stmaryrd,centernot,color,url,amssymb}
\usepackage{amsmath}

\begin{document}

\begin{frontmatter}


\title{On the elastic anatomy of heterogeneous fractures in rock}

\author[UCB]{Fatemeh Pourahmadian}
\author[UMN]{Bojan B. Guzina\corref{cor1}} 

\address[UCB]{Department of Civil, Environmental and Architectural Engineering, University of Colorado, Boulder}
\address[UMN]{Department of Civil, Environmental and Geo- Engineering, University of Minnesota, Twin Cities}

\cortext[cor1]{Corresponding author: tel. 612-626-0789, email {\tt guzin001@umn.edu}}

\date{\today}

\begin{abstract}

\noindent This study examines the feasibility of the full-field ultrasonic characterization of fractures in rock. To this end, a slab-like specimen of granite is subjected to in-plane, O(10$^4$Hz) excitation while monitoring the induced 2D wavefield by a Scanning Laser Doppler Vibrometer (SLDV) with sub-centimeter spatial resolution. Upon suitable filtering and interpolation, the observed wavefield is verified to conform with the plane-stress approximation and used to: (i) compute the maps of elastic modulus in the specimen (before and after fracturing) via a rudimentary application of the principle of elastography; (ii) reconstruct the fracture geometry; (iii) expose the fracture's primal (traction-displacement jump) contact behavior, and (iv) identify its profiles of shear and normal specific stiffness. Through the use of full-field ultrasonic data, the approach provides an unobscured, high-resolution insight into the fracture's contact behavior, foreshadowing in-depth laboratory exploration of interdependencies between the fracture geometry, aperture, interphase properties, and its seismic characteristics. 

\end{abstract}

\begin{keyword}
seismic behavior of fractures,  heterogeneous specific stiffness, full-field ultrasonic sensing
\end{keyword}

\end{frontmatter}

\section{Introduction} \label{sec1}

\noindent Geometric and interfacial properties of fractures and related features (e.g. faults) in rock and other like materials are the subject of critical importance to a wide spectrum of scientific and technological facets of our society including energy production from natural gas and geothermal resources~\citep{Verdon2013b,Verdon2013,Taron2014}, seismology~\citep{McLaskey2012}, hydrogeology~\citep{Cook1992}, environmental protection~\citep{Place2014}, and mining~\citep{Gu1993}. Unfortunately, a direct access to fracture surfaces in rock is, in most field situations, either non-existent or extremely limited (e.g. via isolated boreholes, shafts, or adits), which necessitate the use of remote sensing techniques where the contact law at the boundary of rock discontinuities is often assumed to be linear and represented in a  parametric fashion via e.g. the so-called (shear and normal) specific stiffness, relating the contact traction to the jump in displacements across the interface~\citep{Schoenberg1980}. Despite its heuristic and simplistic nature, the fracture's interfacial stiffness matrix not only is proven to be immediately relevant to the stress and thus stability analyses in rock masses~\citep{Eber2004}, but also bears an intimate connection to the fracture's hydraulic properties~\citep{Pyr1987,pyrak2000,pyrak2016}, and may serve as a precursor of progressive shear failure along rock discontinuities~\citep{Hedayat2014}. In this spirit, the aim of this work is to: 1) non-parametrically expose the true contact law and its spatiotemporal variations along the surface of stationary and propagating fractures in rock, and 2) extract the linearized contact properties in terms of the shear and normal specific stiffness together with their heterogeneous distribution along the fracture. This is accomplished in a laboratory setting by monitoring the full-field interaction of ultrasonic shear waves (propagating through granite specimens) with stationary and advancing fractures via a recently acquired 3D Scanning Laser Doppler Vibrometer (SLDV) that is capable of monitoring triaxial particle velocity, with frequencies up to 1MHz, over the surface of rock specimens with 0.1mm spatial resolution and O(nm) displacement accuracy. Looking forward, the full-field seismic observations such as those presented herein may not only help decipher the true physics of a fracture interface and shine light on fidelity of  classical interface models, but may also provide the ground truth toward validating the next generation of seismic imaging tools for simultaneous reconstruction and interfacial characterization of fractures in rock from remote sensory data~\citep{Fatemeh2015,Fatemeh2017, Fatemeh2017(2)}. 

\section{Experimental program} \label{exp_set}

\noindent The experiments are performed on a slab-like prismatic specimen of charcoal granite with dimensions $0.96$m $\times 0.3$m $\times 0.03$m, mass density $\rho\!=\!2750$kg/m$^3$, nominal Poisson's ratio $\nu\!=\!0.23$, and nominal Young's modulus $E\!=\!62.6$GPa. The latter is obtained from the quasi-static stress-strain curve, measured during uniaxial compression of a cylindrical sample of the same material. 

\begin{figure}[!h] \vspace*{-0mm}
\center\includegraphics[width=0.60\linewidth]{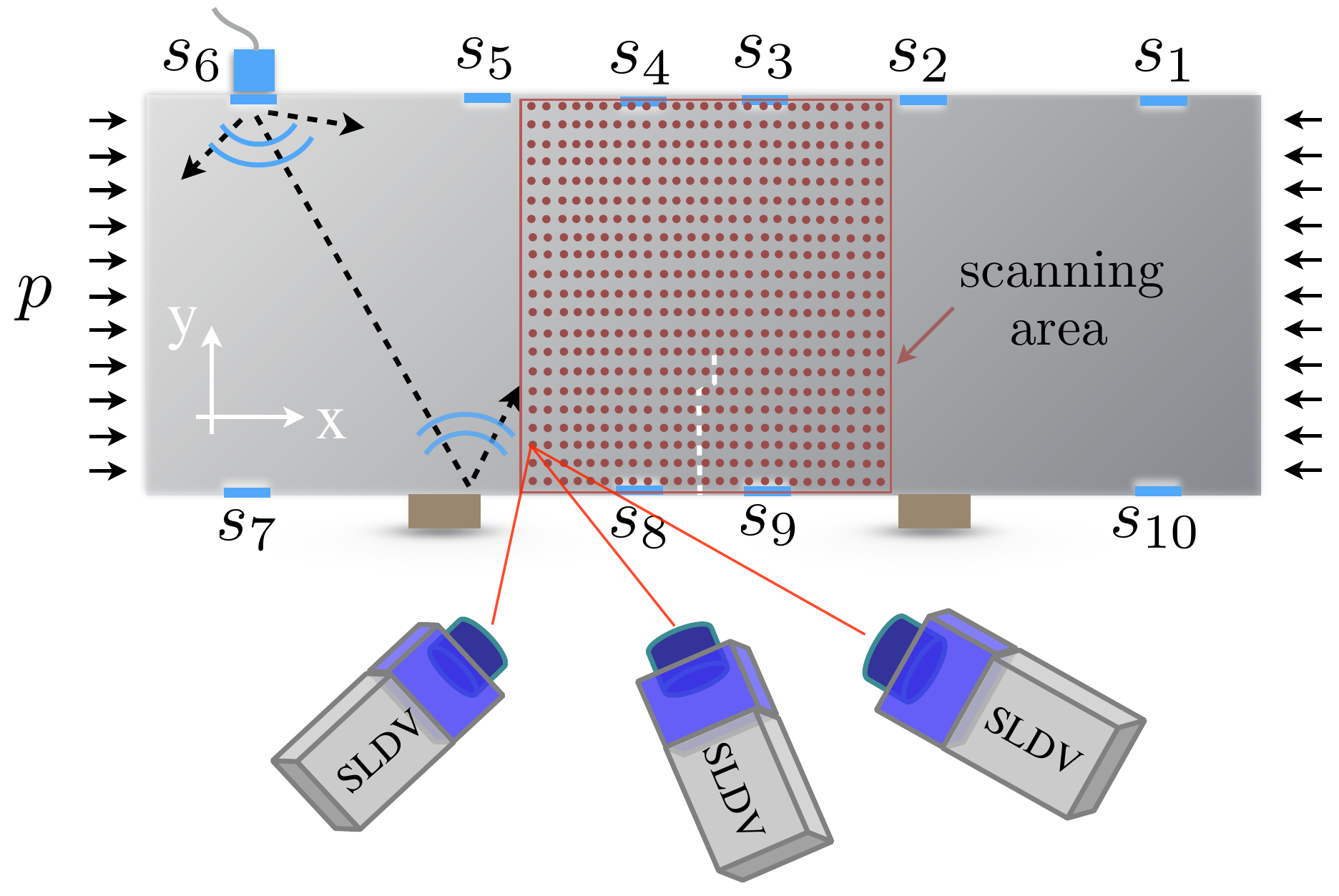} \vspace{2mm}
\center\includegraphics[width=1.0\linewidth]{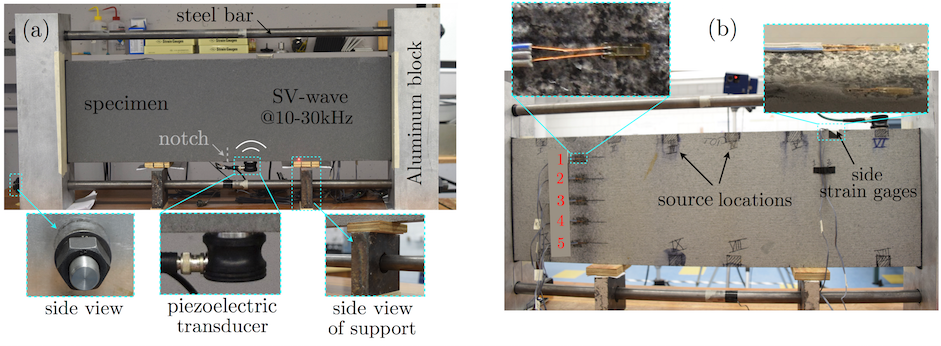} \vspace{-6mm}
\caption{Top panel: full-field ultrasonic monitoring of intact and fractured rock where (i) the specimen is subjected to axial compression $p$; (ii) shear waves are generated by a piezoelectric source at~$s_i$ \mbox{$(i\!=\!1,2,\ldots,10)$}, and (iii) in-plane motion monitored by~SLDV over the scanning area~$\Pi$. Bottom panels: (a)~granite slab in the compression fixture, and (b) specimen's backside instrumented by strain gages.} \label{Exp-sch}
\end{figure}

\paragraph*{Ultrasonic excitation and waveform sensing.} With reference to Fig.~\ref{Exp-sch}, the sample is insonified by shear waves at each of the designated source locations ($s_1$ trhough~$s_{10}$) by an Olympus V151-RB piezoelectric transducer excited by a five-cycle wavelet 

\begin{equation}\label{wavelet}
H(f_{\!c}t) \, H(5\!-\!f_{\!c}t) \, \sin\big(0.2 \pi f_{\!c} t\big) \, \sin\big(2 \pi f_{\!c} t\big), \qquad f_{\!c}\!=\!30\mbox{kHz}
\end{equation}

 where $f_{\!c}$ denotes the center frequency, and $H$ is the Heaviside step function. For each ultrasonic excitation, the \emph{in-plane motion} of a specimen is monitored, using 3D Scanning Laser Doppler Vibrometer (SLDV), over the $32.8$cm $\times\;29.3$cm scanning area~$\Pi$ shown in Fig.~\ref{Exp-sch}. To cater for the full-field analysis of waveforms with wavelengths on the order of 10cm, the ultrasonic motion is captured over a uniform grid of $49\times 37$ measurement locations, providing the spatial resolution of under 8mm in both $x$- and $y$-directions. The featured SLDV system PSV-400-3D by Polytec, Inc. (frequency range DC-1MHz) is capable of capturing the normal and in-plane particle velocity fields with the measurement (resp. spatial) resolution better than 1$\mu${m}/s (resp. 0.1mm), and allows for displacement waveform sensing in the nanometer range~\citep{Polytec}. 

\paragraph*{Two-dimensional wave propagation framework.} For an in-depth analysis of the ultrasonic fracture behavior, the observed in-plane wavefields are interpreted in the context of the \emph{plane stress} approximation~\citep{Mal1969} -- catering for the elastic analysis of thin slabs -- by which the wave motion is constant through the thickness of a specimen. In this case, the \emph{effective} (plane-stress) Poisson's ratio and Young's modulus are computed respectively as $\nu'\!=\!\nu/(1+\nu)$ and $E'\!=\!E(1-\nu'^2)$~\citep{Mal1969}, giving the shear (S-) and compressional (P-) wave velocities in the granite slab as 

\begin{equation}\label{cps}
c_s ~=~ \sqrt{\frac{E}{2(1+\nu)\rho}} ~=~ 3041 \, \, \mbox{m/s}, \qquad  c_p ~=~ \sqrt{\frac{E}{(1-\nu^2)\rho}} ~=~ 4901 \, \, \mbox{m/s}. 
\end{equation}

Accordingly, the shear wavelength $\lambda_s$ in the specimen at 30kHz is approximately $10$cm, giving the shear-wavelenghth-to-specimen-thickness ratio $\lambda_s/h\!\gtrsim\!3.3$. In this range, the phase error committed by the plane stress approximation is about $3\%$~\citep{Lamb1917}. For completenes, an in-depth verification of the plane stress condition, which makes use of the full-field waveform data, is provided in the sequel. 

\begin{figure}[!hbp] \vspace*{-0mm}
\center\includegraphics[width=0.70\linewidth]{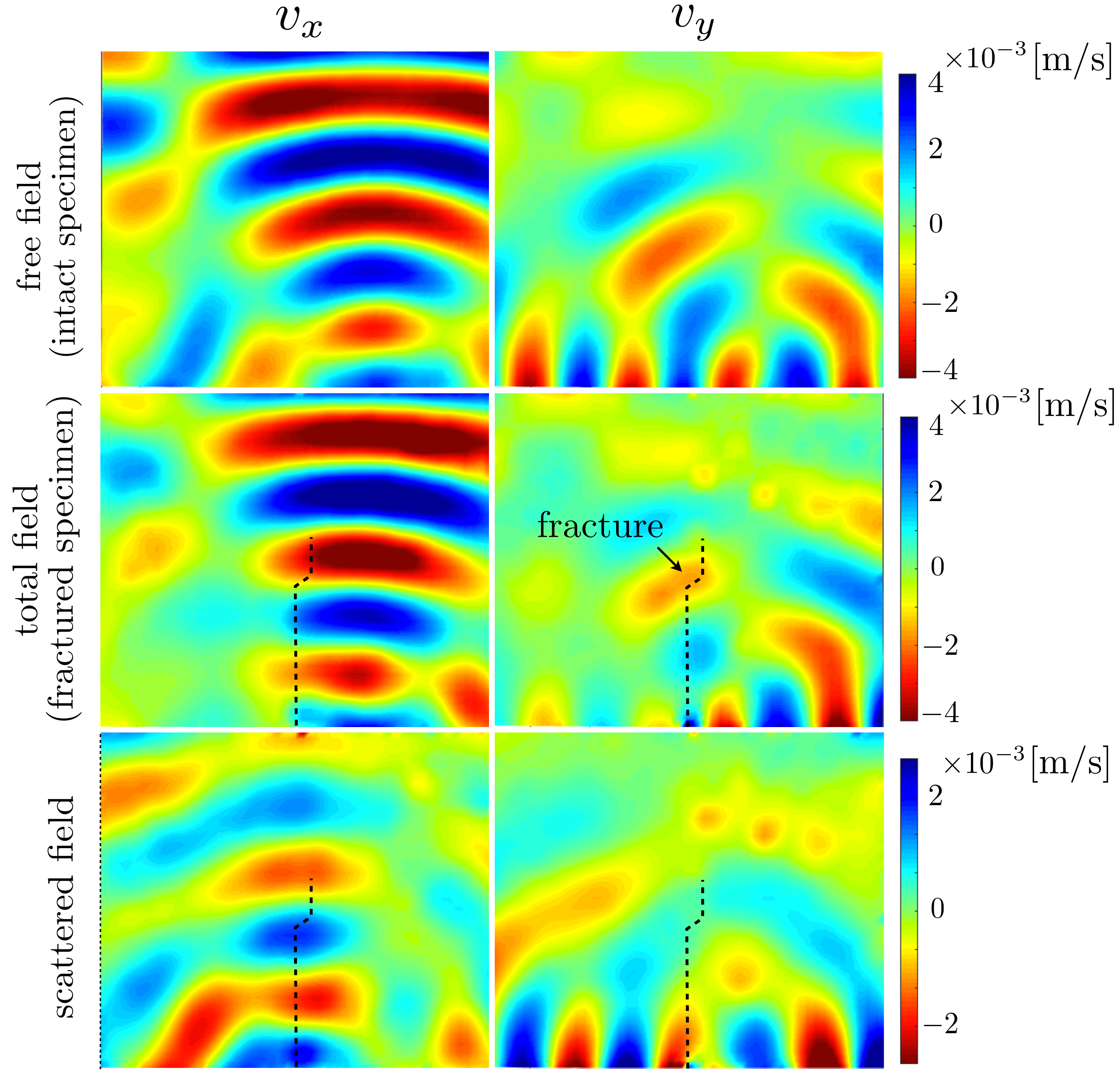} \vspace{-0mm}
\caption{2D wave motion captured in the rock specimen before and after fracturing: snapshot in time of the particle velocity fields $v_x$ and~$v_y$ in the intact sample due to excitation at $s_9$ (top panels); corresponding wavefields in the fractured specimen (middle panels), scattered particle velocity wavefield, computed by subtracting the motion in the intact specimen from that in the fractured specimen (bottom panels).} \label{fr-tot-sc}
\end{figure}

\paragraph*{Step~1: ultrasonic testing of intact rock.} The ultrasonic experiments are first performed on the intact granite slab as in Fig.~\ref{Exp-sch} with no compressive prestress ($p\!=\!0$). Thus obtained measurements furnish the ``baseline'' wave motion in the slab, see the top panels in Fig.~\ref{fr-tot-sc}, that is later used to: (a) expose the intrinsic heterogeneity of the rock's elastic properties prior to fracturing, and (b) directly compute the \emph{scattered wavefield} due to \emph{fracturing-induced} topological and material changes in the specimen.

\paragraph*{Step~2: rock fracturing.} To help fracture the specimen in a controlled fashion, a $4$cm-long notch is manufactured in the middle of its bottom edge as shown in Fig.~\ref{Exp-sch}(a). The granite slab is then fractured in the three-point-bending (3PB) configuration via a 1000kN MTS load frame, where the crack propagation is controlled by a closed-loop, servo-hydraulic system using the crack mouth opening displacement (CMOD) as the feedback signal. The loading process is continued up to approximately $65\%$ of the maximum force in the post-peak regime. Upon completion of the fracturing process, the specimen is unloaded and transferred to the compression frame (see Fig.~\ref{Exp-sch} (a)) for ultrasonic testing.  

\paragraph*{Step~3: ultrasonic testing of fractured rock.}  The ultrasonic experiments are performed using the same setup as in \emph{Step~1} in terms of transducer locations, illuminating wavelet, and scanning area (see Fig.~\ref{Exp-sch}). To investigate the pressure-dependent mechanical response of the macroscopic and microscopic fractures generated in \emph{Step~2}, however, the experiments are performed at both unstressed ($p\!=\!0$ as in \emph{Step~1}) and axially prestressed ($p\!=\!1.5$MPa) states. As shown in Fig.~\ref{Exp-sch}(a), the axial stress $p$ is applied to the specimen via a load frame, composed of two aluminum blocks connected by steel bars threaded on both ends -- which allows for the application of normal compression by screwing the nuts on these threaded portions against the blocks. Thus applied compressive stress and its distribution along the specimen's height is then quantified by using the previously identified (nominal) Young's modulus and five strain-gage readings taken on the back face of the specimen as in Fig.~\ref{Exp-sch}(b). Furthermore, the variations of $p$ through the thickness of a slab -- due to e.g.~out-of-plane moments triggered by the misalignments of the compression frame, are monitored via a pair of longitudinal strain gages placed on the specimen's edge, one next to each face (see Fig.~\ref{Exp-sch}(b)). It is observed that such out-of-plane variations are less than 10\% of the average $p$-value.     

To illustrate the acquired full-field SLDV measurements, Fig.~\ref{fr-tot-sc} plots a snapshot in time of the $v_x$- and $v_y$-particle velocity distributions over the scanning area for the intact specimen (top row), fractured specimen (middle row), and the difference between the two (bottom row) -- signifying the scattered field due to 3PB-induced damage in the sample. In particular, one may note that (for the ultrasonic source at~$s_9$) the surface-breaking notch is a major generator of Rayleigh waves along the bottom edge, while the wavefield scattered by the fracture is polarized primarily in the $x$-direction. 

\section{Full-field ultrasonic characterization} \label{ffuc}

\begin{figure}[!h]
\center\includegraphics[width=0.84\linewidth]{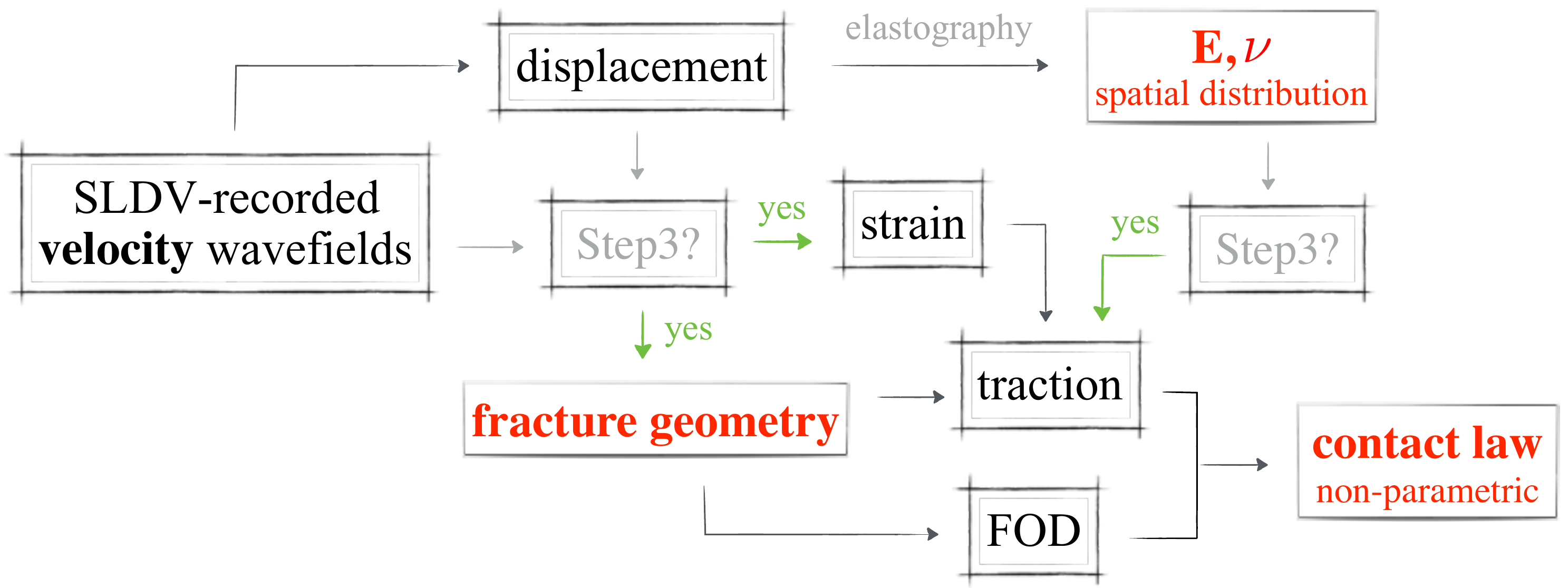} \vspace{-2mm}
\caption{Roadmap for SLDV data processing deployed to: (a)~reconstruct the spatial distribution of rock's elastic properties, and (b)~identify the fracture geometry and its heterogeneous contact law.} \label{chart_exp} \vspace*{-0mm}
\end{figure}

\noindent To expose the elastic fingerprint of fracturing-induced damage in rock, the ultrasonic wavefields captured in \emph{Step~1} and \emph{Step~3} are processed following the scheme outlined in Fig.~\ref{chart_exp}, which entails: (i)~recovery of the spatial map of the Young's modulus in granite specimen before and after fracturing; (ii)~geometric reconstruction of the induced macroscopic fracture, and (iii)~non-parametric recovery the fracture's contact behavior which yields, via Fourier analysis, high-fidelity maps of its complex-valued specific stiffness~\citep{Schoenberg1980} in both normal and shear directions.  

\subsection{Signal processing}\label{DSP}
\begin{figure}[!h]
\center\includegraphics[width=0.59\linewidth]{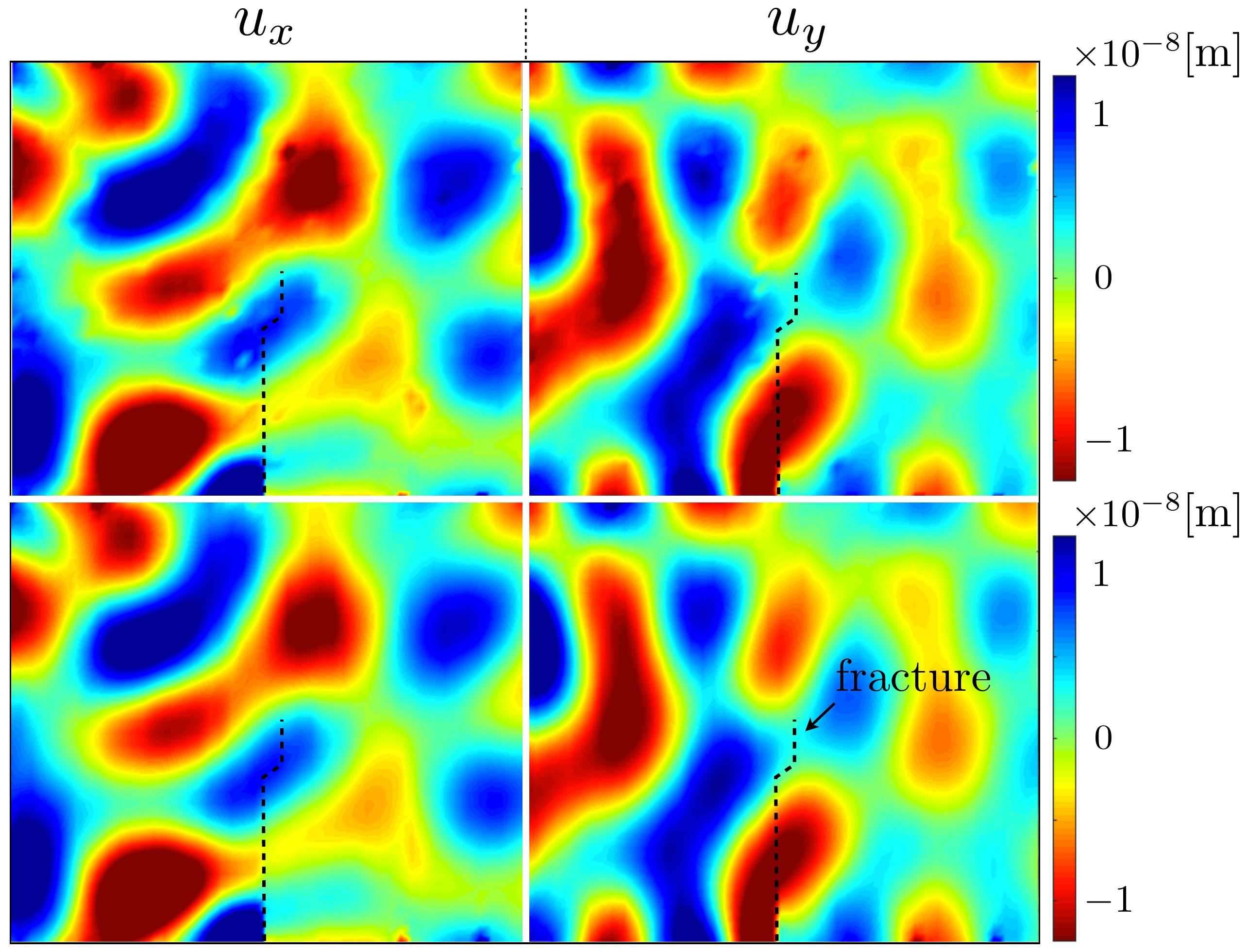} \vspace{-0mm}
\caption{Temporal snapshot of the displacement wavefield ($u_x$, $u_y$) in the fractured specimen,  computed from the SLDV data due to excitation at~$s_9$: ``raw'' displacements waveforms (top), and their smooth counterparts obtained by median-mean filtering and Fourier series representation (bottom).} \label{sig30}
\end{figure}

\noindent  With the above goals in mind, the first step of the signal processing scheme is a recovery of piecewise-differentiable \emph{displacement wavefields} from the SLDV particle velocity records.  To this end, a band-pass filter is designed catering for the frequency spectrum of the excitation wavelet~(\ref{wavelet}) with the center frequency of~30kHz, which is then applied to the particle-velocity records at every scan point. Thus filtered velocity signals, featuring temporal smoothness and differentiability, are subject to numerical integration, resulting in the corresponding displacement fields. The latter waveforms are, however, contaminated by a low-frequency drift i.e.~integration constant, that is eliminated by a high-pass filter with the cut-off frequency at $500$Hz. In this way, one finds the ``primary" displacement fields (illustrated in the top panels of Fig.~\ref{sig30}) which are smooth and differentiable temporally, but not spatially. This calls for further processing, since the featured displacement fields must be differentiated twice in space toward the recovery of the rock's (spatially-dependent) elastic parameters by an application of the medical imaging technique known as \emph{elastography}~\citep{Muth1995}, see Section~\ref{ELS}. To address the problem, a three-step smoothing scheme is next applied to the ``primary'' displacement field at every snapshot in time. First, a median filter with the $\lambda_s/8 \times \lambda_s/8$ window is deployed to eliminate the scan points with exceptionally low signal-to-noise ratio -- which are observable as sudden spikes in the observed waveforms (see Fig.~\ref{sig30}). Second, the resulting wavefields are subject to a moving average filter with the same window size, which has proven to be an effective stabilizer for the next step that involves interpolation. Third, the resulting waveforms are approximated independently in the $x$- and $y$-directions via one-dimensional Fourier series, by including up to seven harmonics. This gives rise to the smooth displacement fields -- illustrated in the bottom panels of Fig.~\ref{sig30}, which are two-times differentiable both in time and space, as required for elastography applications.

One should note, however, that the particle velocity (and thus displacement) fields captured in \emph{Step~3} feature an \emph{intrinsic discontinuity} due to the presence of a macroscopic fracture within the scanning area. Accordingly, the three-step displacement smoothing scheme can be applied only \emph{after} the geometric reconstruction of an apparent fracture, see Section~\ref{GEO}. In this setting the reconstructed fracture boundary, observable as the wavefield discontinuity, is first extended to \emph{split} the scanning window (see the top-left panel of Fig.~\ref{sig30}) so that each ``half'' of the scanning area covers a continuous wavefield that is amenable to the smoothing treatment.

\subsection{Geometric reconstruction}\label{GEO}

\begin{figure}[!bp]
\center\includegraphics[width=0.82\linewidth]{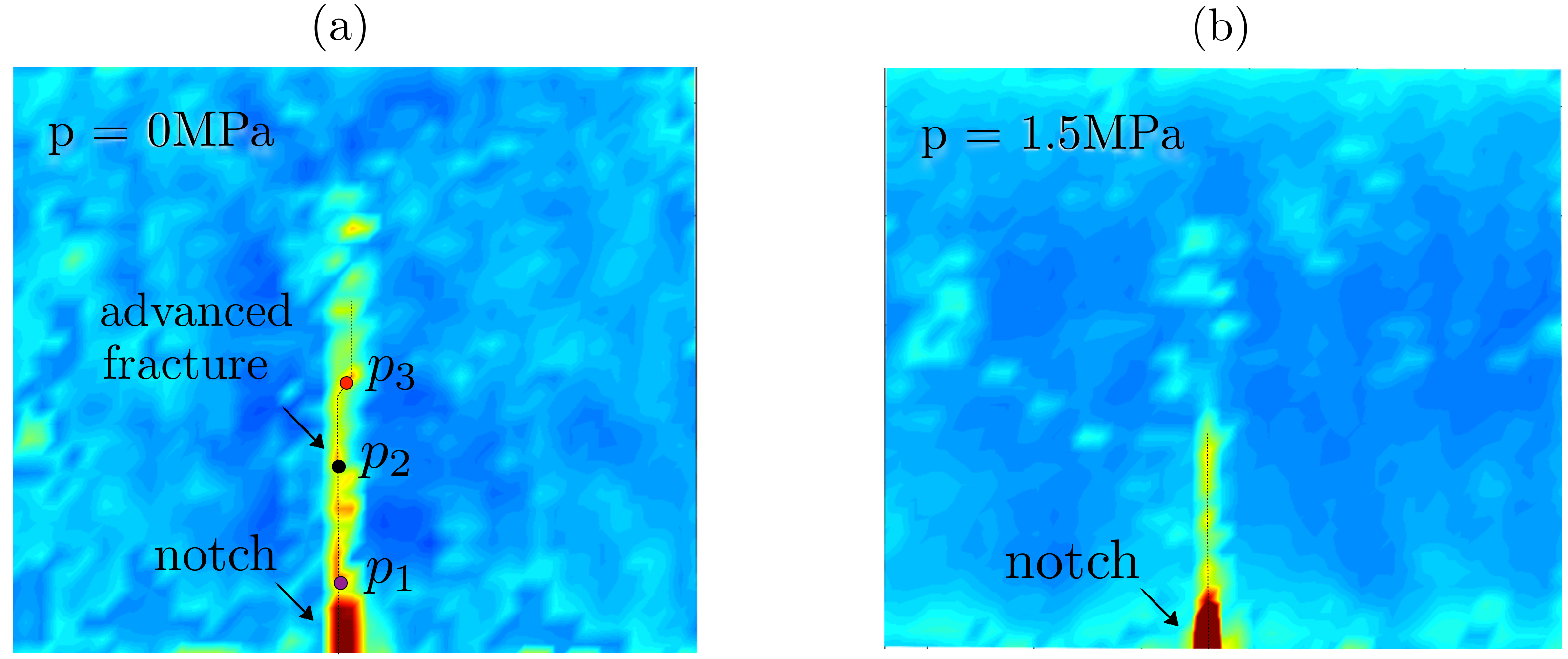} \vspace{-3mm}
\caption{Reconstructed fracture geometry, as exposed by the map of cumulative jump in the SLDV-captured particle velocity field: discontinuity map at (a)~no normal prestress \mbox{($p\!=\!0$)}, and (b) normal prestress $p\!=\!1.5$MPa.} \label{geomtry_recons}
\end{figure}

\noindent It should be noted that the macroscopic fracture in the granite specimen -- induced via 3PB in \emph{Step~2} -- is not visible to the naked eye, necessitating the use of the SLDV records for its geometric reconstruction. In this spirit, one may observe from Figs.~\ref{fr-tot-sc} and~\ref{sig30} that the interaction of ultrasonic waves with the macroscopic fracture gives rise to a discontinuity in the wavefield across its interface. This remark inspires the ensuing reconstruction scheme where the gradient of the particle velocity fields ($v_x$ and $v_y$) is first computed by finite differences at each scan point and every snapshot in time, whose $L^1$-norm is then integrated temporally over the period of observation. Thus computed gradient fields associated with every source location ($s_1$ through~$s_{10}$ in~Fig.~\ref{Exp-sch}) are further superimposed. The resulting discontinuity maps are shown in Fig.~\ref{geomtry_recons}, where the loci of the scan points with extreme jump values clearly expose the support of a hidden fracture.    

\subsection{Elastography}\label{ELS}

\noindent The next aim is to provide a \emph{quantitative map} of the elastic-property perturbations induced in the specimen by fracturing. This is accomplished by invoking the full-wavefield sensing approach known as \emph{elastography}~\citep{Muth1995}, developed for the medical diagnosis of soft tissues, to reconstruct the spatial distribution of elastic moduli in the granite specimen from the full-field SLDV measurements in \emph{Step~1} (resp.~\emph{Step~3}), before (resp.~after) fracturing the sample. The basic idea of elastography, adapted for the purpose of this study, reads: given twice-differentiable displacement fields $u_x$ and $u_y$ over the scanning region (e.g. the bottom panels in Fig.~\ref{sig30}) during certain time span, the distribution of elastic parameters in this area can be computed by solving the 2D Navier equations of motion (i.e. the balance of linear momentum) in terms of the sought-for material properties. More specifically, under the \emph{plane stress} assumption~\citep{Mal1969}, the 2D Navier equations can be conveniently recast as

\begin{equation} \label{elast}
\left[\begin{array}{cc}
\ddot{u}_x & \frac{1}{2}(u_{x,yy} + u_{y,xy}) - u_{y,yx} \\*[2 mm] \ddot u_y & \frac{1}{2}(u_{y,xx} + u_{x,yx}) - u_{x,xy} 
\end{array}\right] \;
\left[\begin{array}{c} c_p^{-2} \\*[2 mm] \nu \end{array}\right] ~=~ 
\left[\begin{array}{c} \frac{1}{2}(u_{x,yy} + u_{y,xy}) + u_{x,xx} \\*[2 mm] \frac{1}{2}(u_{y,xx} + u_{x,yx}) + u_{y,yy} 
\end{array}\right],
\end{equation}

where over-dots indicate temporal differentiation; $f_{,\alpha} \!\!=\! \partial{f}/\partial{\alpha} ~(\alpha\!=\!x,y)$, and $c_p\!=\!\sqrt{{E}/{(\rho \tiny(1-\nu^2) \tiny)}}$ is the plane-stress P-wave speed, see~(\ref{cps}). Note that the implicit postulate in~(\ref{elast}) is that both Young's modulus and Poisson's ratio are \emph{locally} constant; otherwise, their spatial derivatives must be included in the vector of unknowns, see~\citep{Barb2004}, which requires additional constraints in order to solve the problem. For completeness, the validity of such simplifying assumption will be discussed shortly. A sensitivity analysis also suggests that for a robust full-field inversion of $c_p(x,y)$ via~\eqref{elast}, the Poisson's ratio should be excluded from the list of  unknowns by adopting its nominal value, $\nu\!=\!0.23$. This is motivated by notable insensitivity of the recovered distribution of~$c_p(x,y)$ to the Poisson's ratio when assuming $\nu\!=\!\mbox{const.}$ in the range~$0.2\!\leqslant\!\nu\!\leqslant\!0.3$. In this setting,~\eqref{elast} is recast as the overdetermined system    

\begin{equation} \label{elast2}
\begin{aligned}
& \mathcal{L}_{\alpha}(u_x,u_y) ~=~ \rho \ddot{u}_{\alpha}, \qquad \alpha \in \{x,y\} \\
& \mathcal{L}_\alpha ~=~ \dfrac{E}{1\!-\!\nu^2} \, \bigg[ \frac{1\!-\!\nu}{2} (u_{\alpha,\beta\beta} + u_{\beta,\alpha\beta}) +  
u_{\alpha,\alpha\alpha} + \nu \hspace*{1pt}u_{\beta,\beta\alpha} \bigg], \qquad \alpha \neq \beta \in \lbrace x,y \rbrace,
\end{aligned}
\end{equation}

that is then transformed to the frequency domain and solved for $E$ at every scan point.  

On denoting the discrete Fourier transform operator by $\mathcal{F}[\cdot](f)$, the inversion of the balance of linear momentum~(\ref{elast2}) for $E$ within the scanning region is implemented over the bandwidth $\Omega_f\!=\!f_{\!c}\!\pm5$kHz. To eliminate the data with poor signal-to-noise ratio, a frequency-dependent weighting factor $\gamma_{\alpha}(f)$ is assigned to each balance equation in~(\ref{elast2}) in the sense that 

\begin{equation}\notag
\left.
\begin{aligned}
\Big| \mathcal{F} \Big[E^{-1}\mathcal{L}_\alpha \Big](f) \Big| &\geqslant 
\tfrac{1}{2} \max_{\beta\in\{x,y\}} \max_{\Omega_f} \Big|\mathcal{F} \Big[E^{-1}\mathcal{L}_\beta\Big] \Big|& \\
\Big| \mathcal{F} [ \rho \ddot u_\alpha](f) \Big| &\geqslant \tfrac{1}{2} \max_{\beta\in\{x,y\}}\max_{\Omega_f} \Big|\mathcal{F}[\rho \ddot u_\beta]\Big|&
\end{aligned} \!\!\right\} \quad \Longrightarrow \quad \gamma_{\alpha}(f) \!=\! 1, \quad \mbox{else}~~ \gamma_{\alpha}(f) \!=\! 0. 
\end{equation}

Such criterion is particularly useful in situations where the wavefield is strongly polarized in either $x$- or $y$-direction, as it eliminates the ``null'' component of the balance equation -- whose consideration would inherently bring about instability to the inverse solution. In this setting the Young's modulus is computed, at each scan point, as a least-squares minimizer of the misfit functional $\sum_{\alpha\in\{x,y\}}\big\| \gamma_\alpha\mathcal{F}[\mathcal{L}_\alpha(u_x,u_y)-\rho\ddot{u}_\alpha]\hspace*{2pt}\big\|_{\Omega_f}^2$. Such modulus map $E(x,y)$, obtained separately for each ultrasonic source location $s_m$ ($m\!=\!1,2,\ldots 10$), is then averaged over~$m$ for further robustness. 
\begin{figure}[!h]
\center\includegraphics[width=0.99\linewidth]{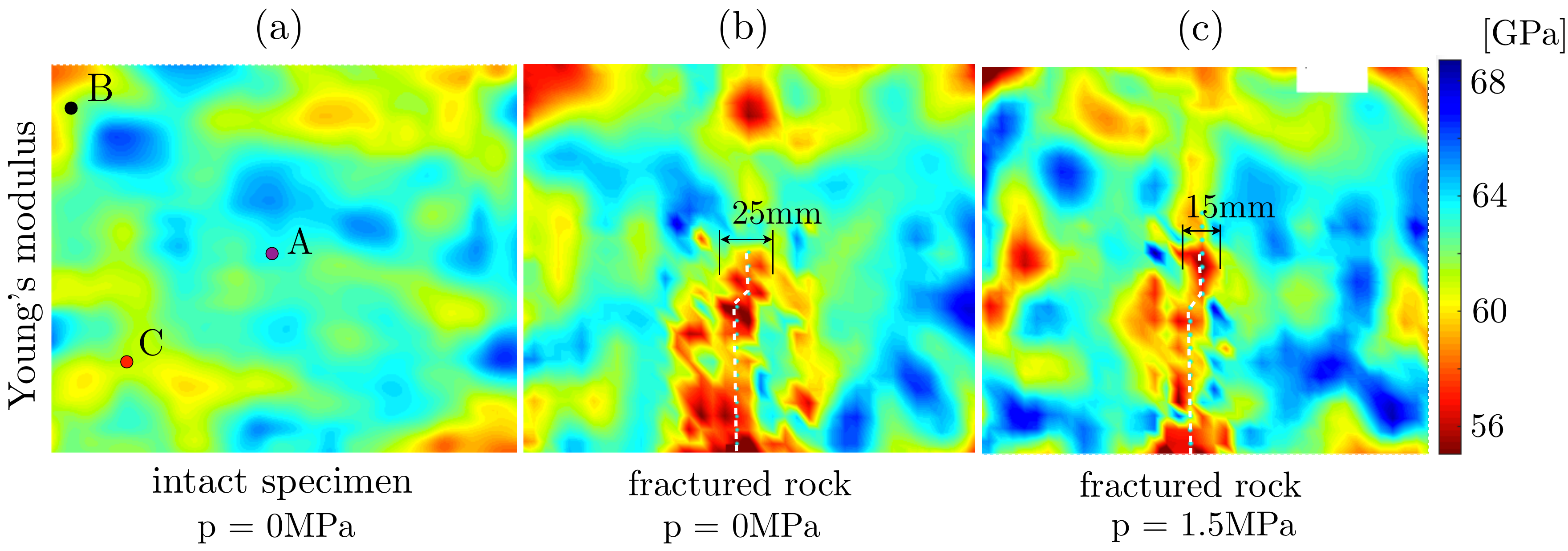} \vspace{3mm}
\center\includegraphics[width=0.55\linewidth]{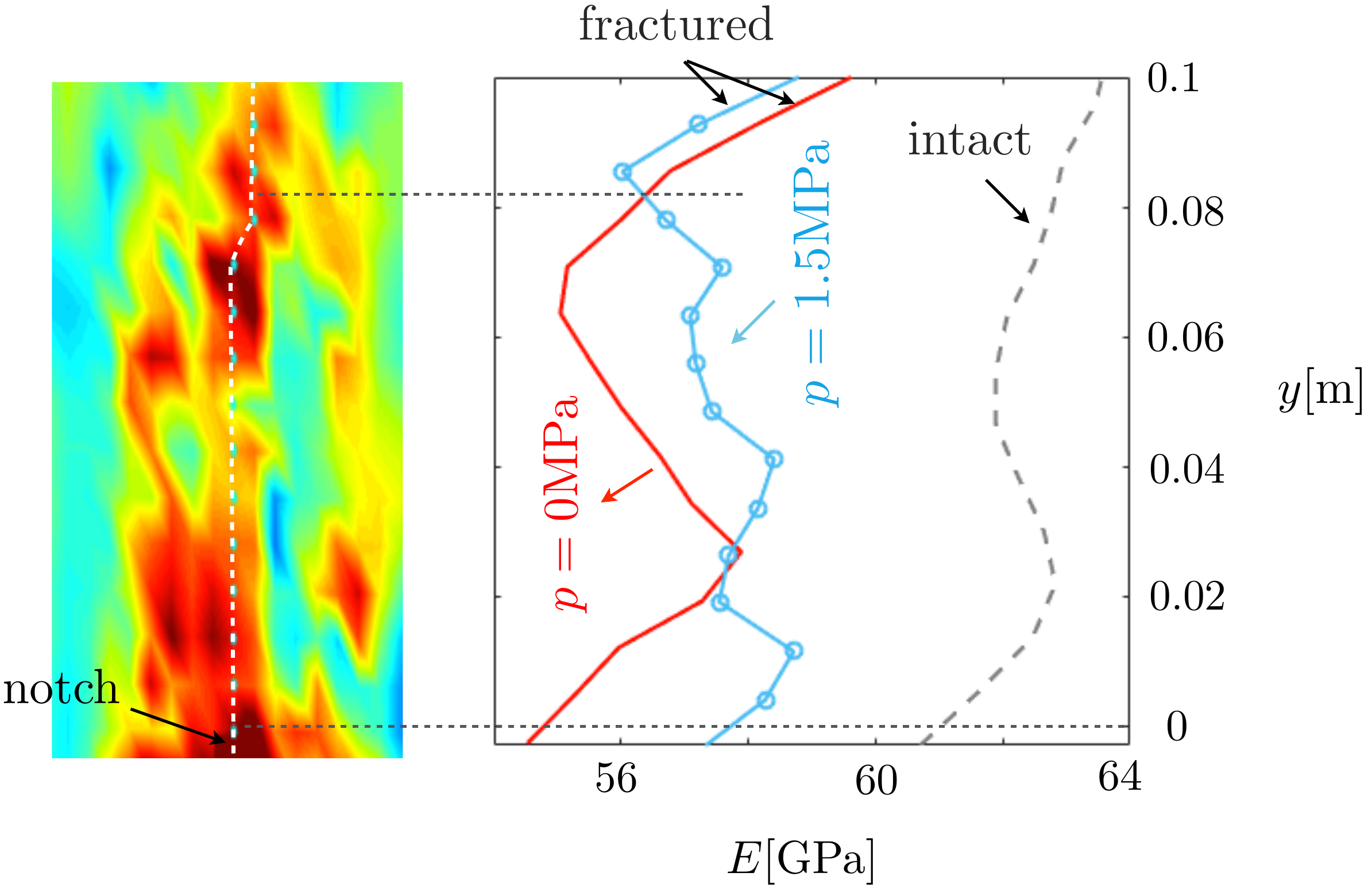} \vspace{-3mm}
\caption{Top panels: map of the Young's modulus in charcoal granite, reconstructed over the scanning area~$\Pi$: (a)~intact specimen, (b)~fractured specimen with no normal prestress (\mbox{$p\!=\!0$MPa}), and (c)~fractured specimen with \mbox{$p\!=\!1.5$MPa}. Bottom panels: 
reconstructed $E$-map in the vicinity of the fracture at \mbox{$p\!=\!0$} (left), and $y$-variation of~$E$, averaged over the width of the respective damage zone at~\mbox{$p\!=\!0$} and~\mbox{$p\!=\!1.5$MPa} (right). The reference ``intact'' profile is obtained by averaging~$E$ in panel~(a) ofer the 15mm-wide strip surrounding the reconstructed fracture.} \label{elastography} \vspace*{-3mm}
\end{figure} 

The evolution of the Young's modulus in rock due to fracturing -- as computed from the SLDV records -- is shown in Fig.~\ref{elastography}, plotting the spatial distribution of $E(x,y)$ in (a)~the intact specimen, (b)~fractured specimen with no axial precompression~($p\!=\!0$), and (c) fractured specimen with $p\!=\!1.5$MPa. On the basis of these results, the \emph{damage zone} -- identified as a neighborhood of the macroscopic fracture where the elastic modulus notably decreases due to microcracking, is characterized by the apparent width of 25mm (resp.~15mm) in the specimen subject to 0MPa (resp.~1.5MPa) normal prestress. This result agrees with the previously reported width of 10-20mm for the \emph{process zone} in charcoal granite, as identified from the acoustic emission records~\citep{Ziet1998}. In support of the shrinking damage zone with increasing~$p$, one may also note that compression tends to close the microcracks which ``cloaks'' the damage against probing ultrasonic waves. One key insight from the featured elastography maps is a \emph{quantitative characterization} of the damage zone in terms of the mean ``damaged'' modulus of 56.8GPa (resp.~57.8GPa) at the uniaxial compression of 0MPa (resp.~1.5MPa) -- obtained by averaging $E(x,y)$ over the strip of width 25mm (resp.~15mm) surrounding the fracture. These ``damaged'' moduli should be compared with the mean ``intact'' value of $E\!=\!61.7$GPa, computed by averaging the $E$-map in Fig.~\ref{elastography}(a). Note that the latter mean value is within 2\% from the nominal Young's modulus, $E\!=\!62.6$GPa, obtained from the uniaxial compression on a cylindrical sample of charcoal granite. For further reference, Fig.~\ref{elastography} also plots the \emph{local variation} in the $y$-direction of the intact and damaged modulus values, averaged (as applicable) over the width of the apparent damage zone.  

\subsection{Verification}\label{Ver}
 
\begin{figure}[!h]
\center\includegraphics[width=0.94\linewidth]{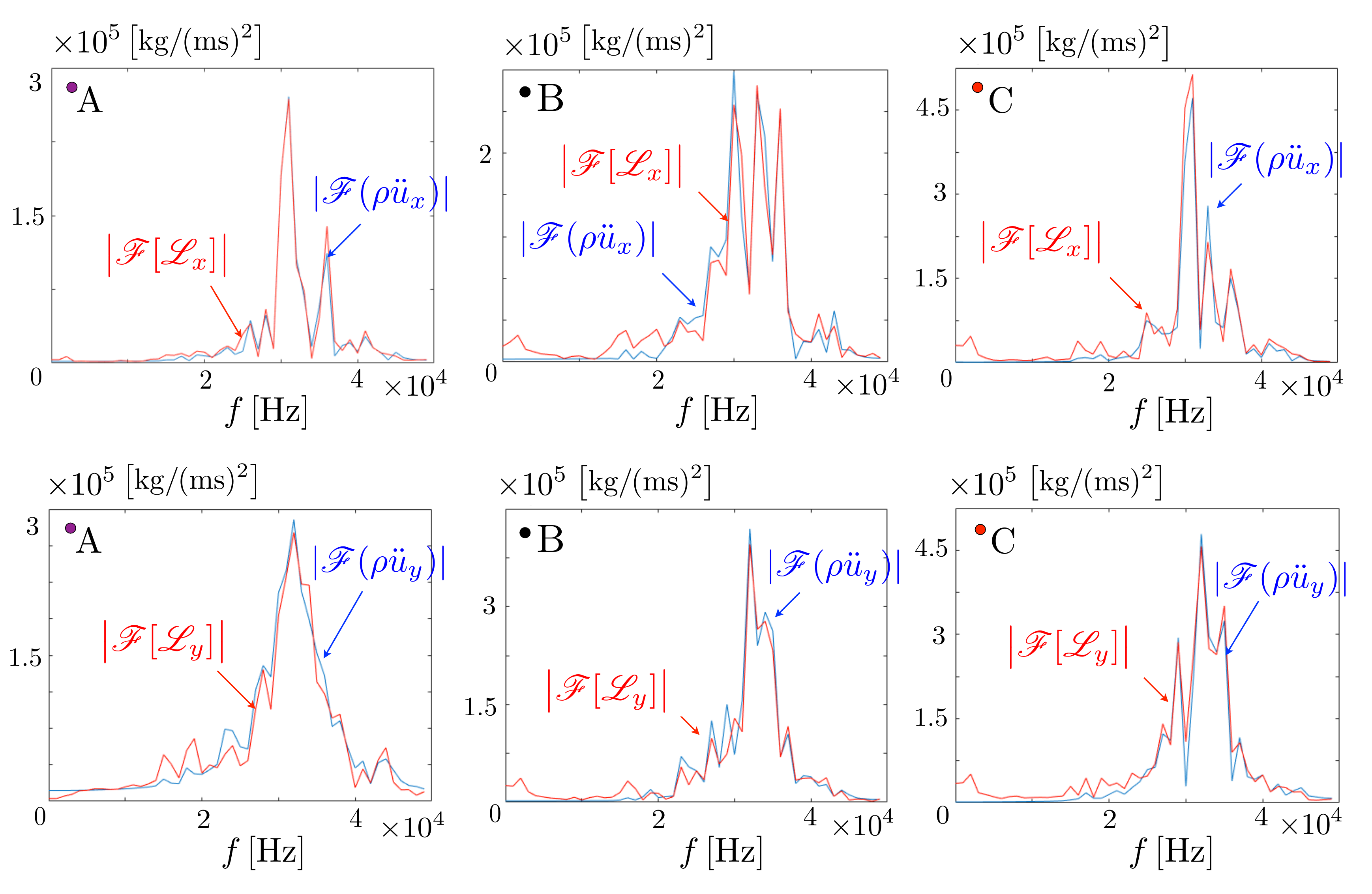} \vspace{-2mm}
\caption{Balance of the plane-stress Navier equations~\eqref{verify} at three scan points (A, B, and C in Fig.~\ref{elastography}(a)) versus frequency. The SLDV measurements are taken in the intact specimen, excited at~$s_9$.} \label{FSNavier}
\end{figure}

\begin{figure}[!h]
\center\includegraphics[width=0.70\linewidth]{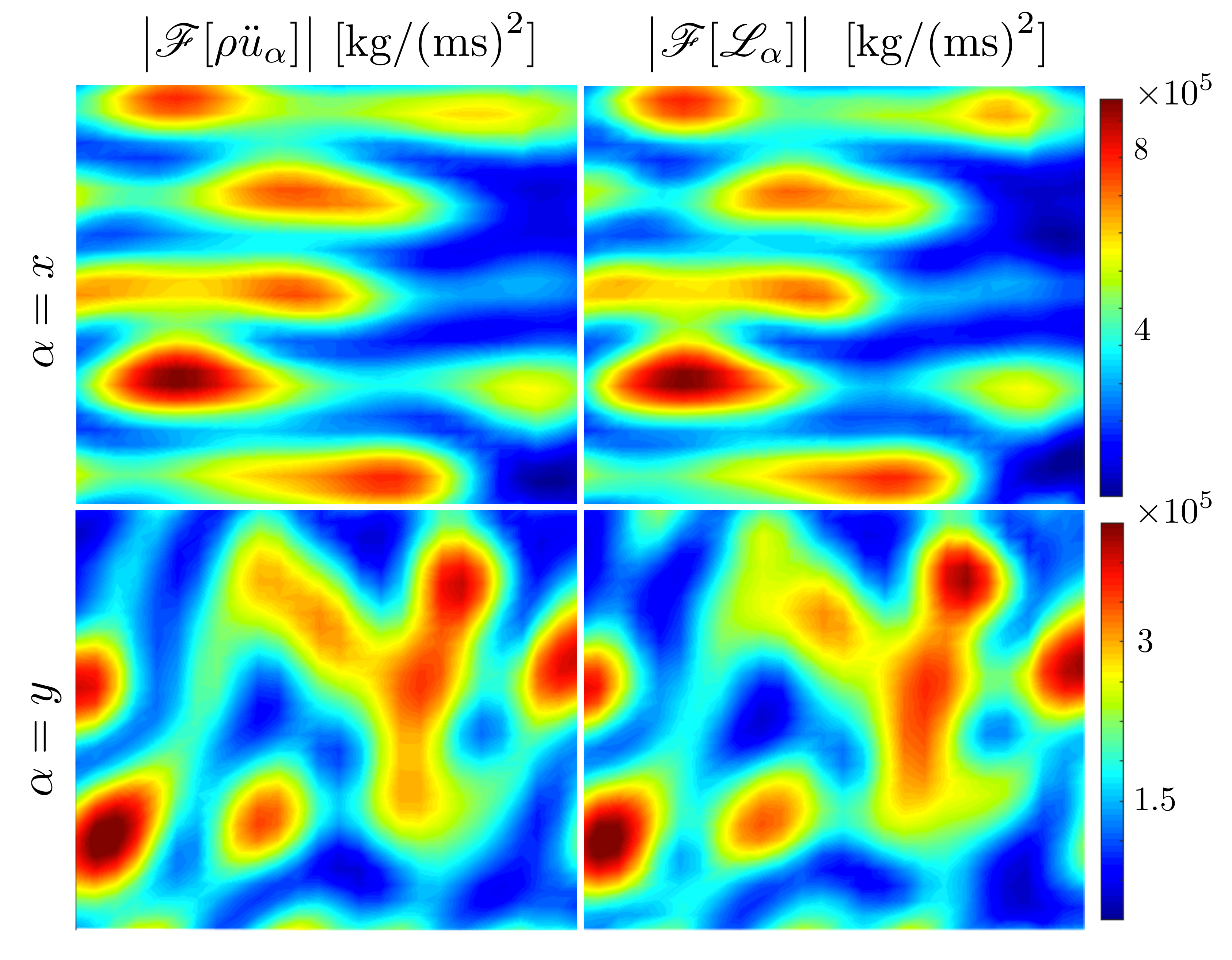} \vspace{-2mm}
\caption{Balance of the plane-stress Navier equations~\eqref{verify} over the scanning area~$\Pi$ at~30kHz. The SLDV measurements are taken in the intact specimen, excited at~$s_9$.} \label{Naviercheck} \vspace{-5mm}
\end{figure}

\noindent Three critical hypotheses underpinning the foregoing analysis state that: (i)~filtered, integrated, and interpolated SLDV records provide accurate representation of the germane ($O(10$nm)) displacement fields; (ii)~wave propagation inside the rock slab can be well approximated by the plane stress model, and (iii)~elastic properties of the rock vary slowly so that their gradients can be neglected in the elastography analysis. Taking advantage of the available full-field measurements, these assumptions can be jointly verified through a close (frequency-domain) examination of the Navier equations of motion, specifically written for the state of plane stress with locally-constant elastic parameters, i.e.   

\begin{equation}\label{verify}
\mathcal{F}[\mathcal{L}_\alpha(u_x,u_y)]=\mathcal{F}[\rho \ddot{u}_\alpha], \qquad \alpha = x,y 
\end{equation}

where $\mathcal{L}_\alpha$ is given by~(\ref{elast2}). Recalling Fig.~\ref{elastography}(a), this balance is first verified at three control points (A, B, and C) within the scan region in Fig.~\ref{FSNavier}(a), which compares $|\mathcal{F}[\mathcal{L}_\alpha(u_x,u_y)|$ and $|\mathcal{F}[\rho \ddot{u}_\alpha]|$ due to source at~$s_9$ with over the relevant frequency range, $[0,50$kHz]. As can be seen from the display, the left- and right-hand sides of the momentum balance equations closely follow each other at all three points, over the entire frequency range, and in both coordinate directions. Next, the veracity of~\eqref{verify} over the entire scanning area is shown in Fig.~\ref{Naviercheck}, again demonstrating remarkable agreement of the model. 

\subsection{Fracture's contact behavior}\label{S}

\begin{figure}[!h]
\center\includegraphics[width=0.8\linewidth]{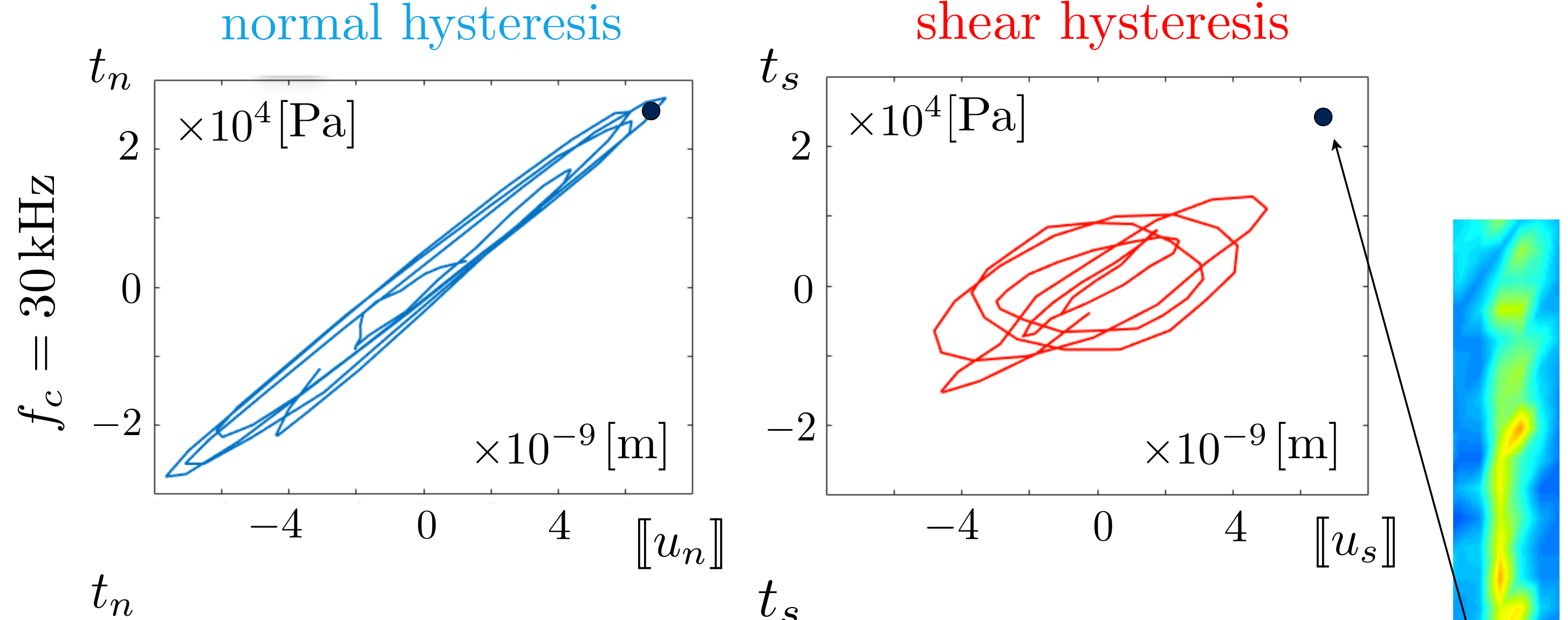} \vspace{-2mm}
\caption{Transient interfacial behavior in the shear $(t_s$ vs. $\llbracket u_s\rrbracket)$ and normal $(t_n$ vs. $\llbracket u_n\rrbracket)$ directions at one point along the fracture with~\mbox{$p\!=\!0$}. The ultrasonic source is located at~$s_9$.} \label{hys} 
\end{figure}

\noindent The above results cater for a high-fidelity estimation of dynamic stresses in a rock specimen which, given the fracture geometry (see Fig.~\ref{geomtry_recons}), enables the computation of shear and normal tractions $(t_s, t_n)$ on the fracture due to action of ultrasonic waves. Similarly, the profiles of shear and normal displacement discontinuity across the fracture $(\llbracket u_s\rrbracket, \llbracket u_n\rrbracket)$, known as the fracture opening displacement (FOD), can be computed as a \emph{jump} between the piecewise-smooth displacement wavefields -- computed separately on \emph{each side} of the fracture as described in Section~\ref{DSP} (see also Fig.~\ref{sig30}). With the germane state variables now in place, one may expose the ``true'' ultrasonic behavior at the fracture interface, without parametrization, by plotting the contact traction versus FOD in both shear and normal directionsat. As an illustration, Fig.~\ref{hys} plots the typical $t_n-\llbracket u_n\rrbracket$ and $t_s-\llbracket u_s\rrbracket$ responses in the time domain, whose hysteretic character reflects the frictional nature of the rock-to-rock contact. It is also seen that, at the featured observation point, the FOD due to ultrasonic excitation at~$s_9$ is primarily normal.

Notwithstanding the possible (weak) presence of nonlinear effects, the contact behavior in Fig.~\ref{hys} is for simplicity described in terms of the \emph{linear-slip} model~\citep{Schoenberg1980}, allowing for a compact description of the fracture's heterogeneous interfacial characteristics in terms of the shear and normal \emph{profiles of specific stiffness}, $k_s$ and $k_n$. Given a rich frequency content of the transient ultrasonic fields, the contact tractions and corresponding FOD components are first transformed to the frequency domain, where the reconstruction of the $k$-profiles is carried out over the bandwidth $\Omega_f\!=\!f_{\!c}\!\pm5$kHz. Similar to the approach adopted for elastography analysis, the evaluation is performed only at frequencies~$f_j\in\Omega_f$ ($j\!=\!1,2,\ldots$) with substantial signal-to-noise ratio, where 

\[
\big|\mathcal{F}[t_\alpha](f_j)| \geqslant 0.4 \max_{\Omega_f}\big|\mathcal{F}[t_\alpha]\big| \quad \mbox{and} \quad
\big|\mathcal{F}\big[\llbracket u_\alpha\rrbracket \big](f_j)| \geqslant 0.4 \max_{\Omega_f}\big|\mathcal{F}\big[\llbracket u_\alpha\rrbracket\big]\big|, \qquad
\alpha\!=\!s,n.
\]

\begin{figure}[!htp]
\center\includegraphics[width=0.94\linewidth]{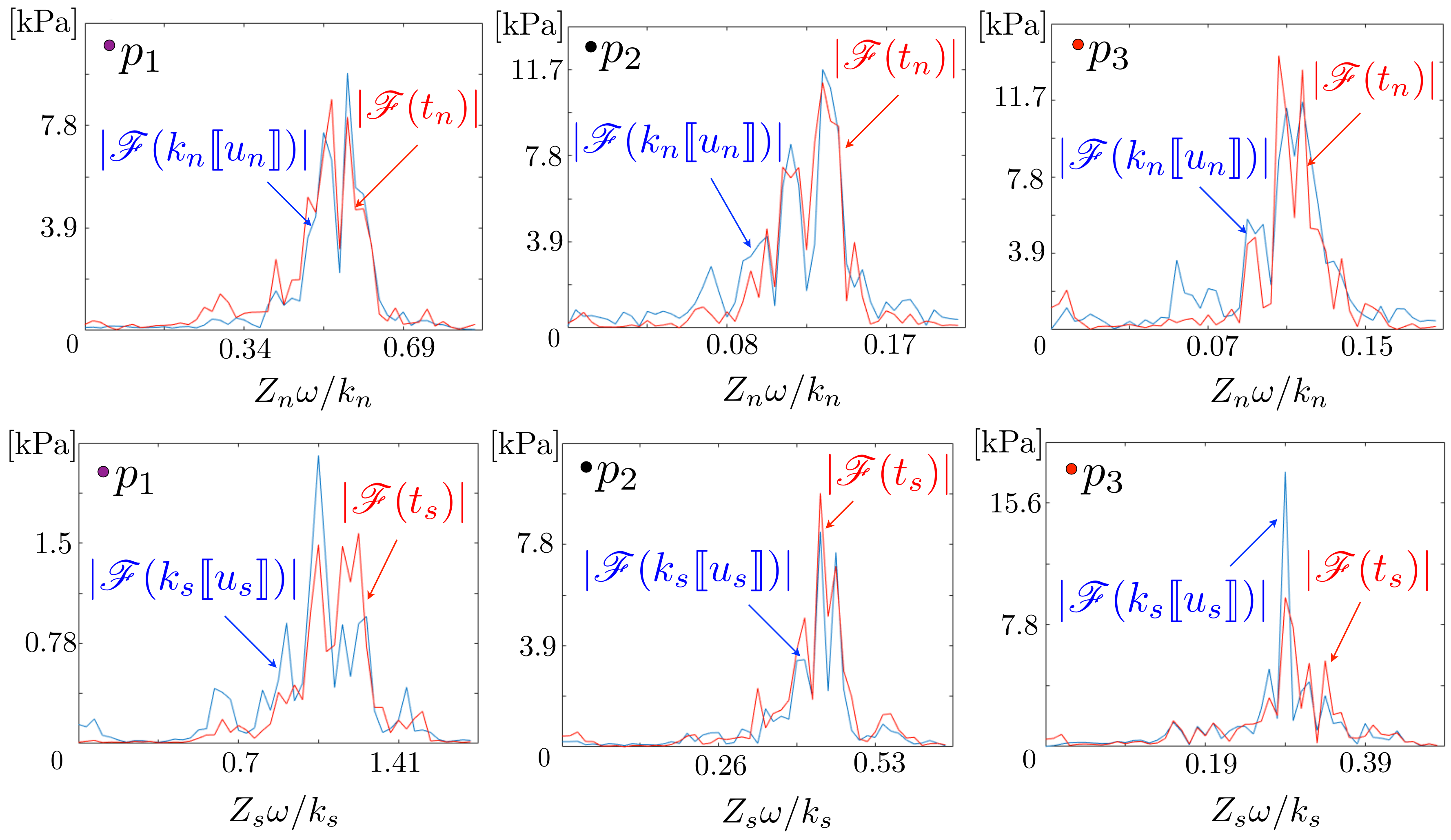} \vspace{-2mm}
\caption{Fidelity of the linear-slip contact condition~\mbox{$\mathcal{F}[t_\alpha]\!=\!\mathcal{F}[k_\alpha \llbracket u_\alpha \rrbracket]$} versus dimensionless frequency at points $p_1, \, p_2$ and $p_3$ in Fig.~\ref{geomtry_recons}(a). The wave motion is generated at~$s_9$ with no normal prestress, i.e. $p\!=\!0$. The range of comparison is [0,50kHz].} \label{LinBal} \vspace*{-00mm}
\end{figure}

 The resulting sets $k_\alpha^{j} = \mathcal{F}[t_\alpha](f_j)/\mathcal{F}\big[\llbracket u_\alpha\rrbracket \big](f_j)$ are then averaged over~$j$ to obtain the \emph{effective complex values}~\citep{Schoenberg1980} of the specific stiffness coefficients, $k_s$ and~$k_n$, at any given point along the fracture. Note that the imaginary parts of $k_s$ and~$k_n$ quantify the effects of frictional dissipation, and can be nominally related to the notion of ``specific viscosity'', found in~\citep{nolt1990} to improve the prediction of the seismic response of fractures in dry rock. Taking the ultrasonic measurements from \emph{Step~3} source at~$s_1$ as an example, Fig.~\ref{LinBal} illustrates the fidelity of the linear slip model with \emph{frequency-independent}~$k_s$ and~$k_n$ by comparing $\big|\mathcal{F}[t_\alpha]\big|$ versus $\big|k_\alpha \mathcal{F}\big[\llbracket u_\alpha\rrbracket \big]\big|$ (at points $p_1, \, p_2$ and $p_3$ in Fig.~\ref{geomtry_recons}(a)) over the bandwith $[0,50$kHz]. For generality, each variation is plotted in terms of the respective dimensionless frequency, $\omega Z_\alpha/k_\alpha$, where $Z_s\!=\!\rho c_s$ and  $Z_n\!=\!\rho c_p$ denote the germane seismic impedances. As can be seen from the display, the use of rate-independent $k_s$ and~$k_n$ yields reasonable description of the ultrasonic contact behavior over a fairly wide bandwidth. In particular, there are no clear trends in terms of the misfit between $\big|\mathcal{F}[t_\alpha]\big|$ and $\big|k_\alpha \mathcal{F}\big[\llbracket u_\alpha\rrbracket \big]\big|$ that would suggest the introduction of frequency-dependent~$k_\alpha$. 

In the context of  Fig.~\ref{LinBal}, it is also worth noting that for normally-incident waves, the use of low-frequency illumination ($\omega Z_\alpha/k_\alpha\!\lesssim\!0.1$) yields the trivial trasmission coefficient $|T|\!\approx\!1$ and maximum group time delay~$t_g$ across the fracture~\citep{nolt1987}, while also resulting in trivial phase velocity of the (slow and fast) interfacial waves, $c_{int}\!\approx\!c_s$~\citep{nolt1996}. In contrast at high frequencies ($\omega Z_\alpha/k_\alpha\!\gg\!1$), the fracture is primarily sensed by the diminishing transmission coefficient $|T|\!\ll\!1$ and the reduced velocity of interfacial waves $c_{int}\!\approx\!c_R$, equaling the Rayleigh wave speed. In this setting, the proposed sensing platform is well suited to deal with a broad spectrum of excitation frequencies -- both ``low'' (as in this study) and ``high'' -- as it holistically considers: (a) body and interfacial waves, (b) amplitude and phase information, and (c) transmitted and reflected components of the incident ultrasonic field. 

\begin{figure}[!h]\vspace{-3mm}
\center\includegraphics[width=0.84\linewidth]{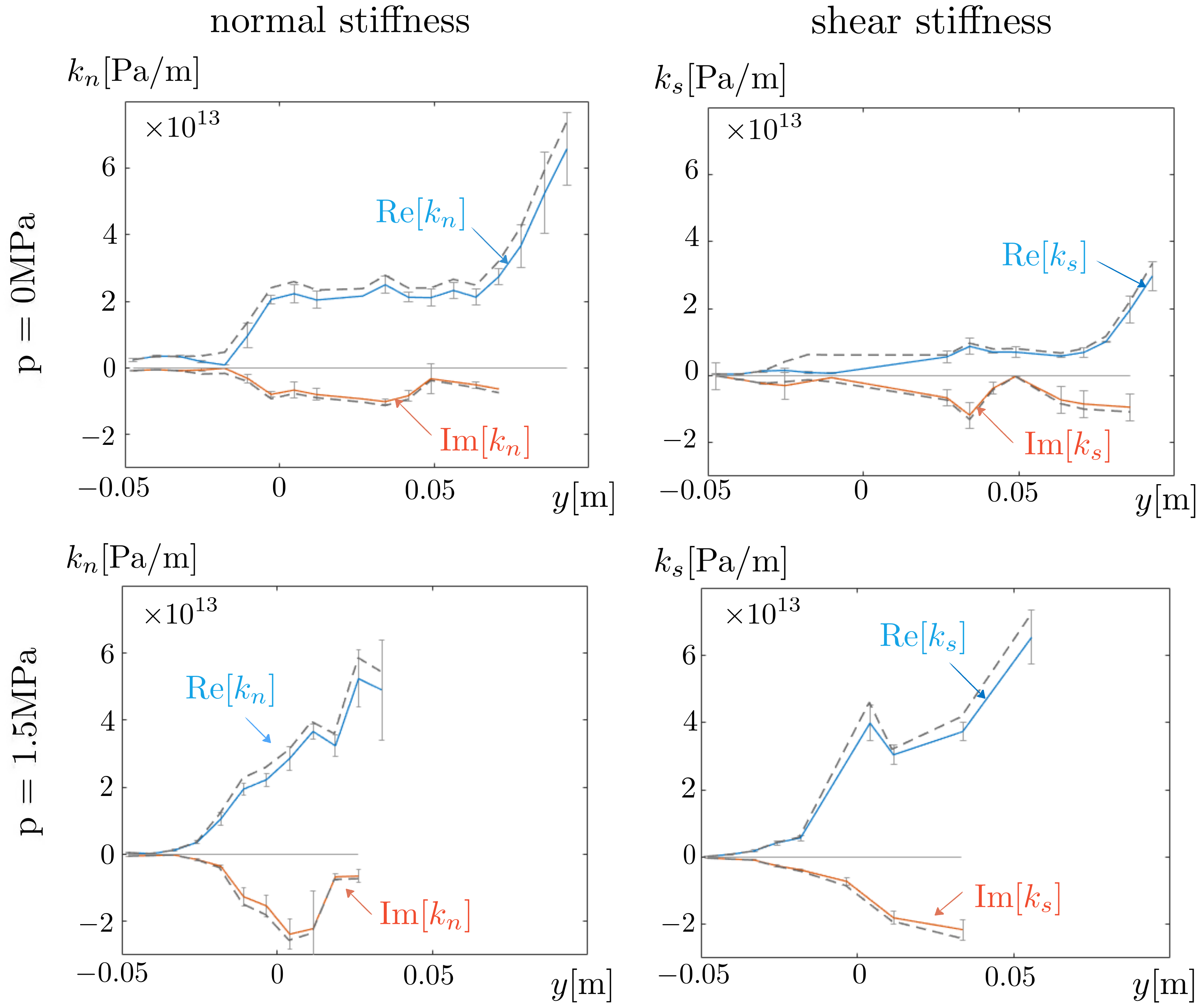} \vspace{-0mm}
\caption{Profiles of specific stiffness along the fracture subjected to normal stress \mbox{$p\!=\!0$} (top panels) and~\mbox{$p\!=\!1.5$MPa} (bottom panels), obtained using \emph{damaged modulus}, $E(y)$, as a means to compute the contact tractions. The real and imaginary profiles of~$k_\alpha$ are given by their mean values (solid lines) and standard deviations (error bars), derived from ten independent reconstructions due to source locations $s_1, s_2, \ldots, s_{10}$.  The $k_\alpha$ profiles computed on the basis of the \emph{nominal modulus} (\mbox{$E\!=\!62.6$GPa}) are indicated by dashed lines.} \label{s30} \vspace*{-4mm}
\end{figure}

The foregoing identification procedure is next applied \emph{independently} to the full-field data stemming from \emph{each} source location $s_m$ $(m\!=\!1,2,\ldots,10)$, allowing for a robust evaluation of~$k_\alpha$ by averaging over~$m$. Thus computed $k_s$ and $k_n$~profiles are shown, in terms of their mean values and standard deviations for $p\!\in\!\{0,1.5$MPa\}, in Fig.~\ref{s30}. To expose the damage-induced perturbations in the contact behavior, specific stiffness profiles are evaluated using both the \emph{damaged modulus} ($E(y)$ in Fig.~\ref{elastography}) and the \emph{nominal modulus} ($E\!=\!62.6$GPa) as a means to compute the contact tractions~$t_\alpha$. From the data in Fig.~\ref{s30}, the following observations are pertinent.  
\begin{itemize}[leftmargin=*] \vspace*{-3mm}
\item The obtained profiles of specific stiffness are on the order of 10TPa/m, which agrees with the results of earlier ultrasonic studies~\cite[e.g.][]{nolt1987, nolt1990,nolt1996}

\item The initial segments of vanishing contact stiffness ($k_s,k_n\!\simeq\!0$) correctly reflect the location of the notch. 

\item The real parts of $k_s$ and $k_n$ increase toward the fracture tip, as expected in light of the diminishing aperture and thus growing contact area in that direction. 
 
\item The distributions of Im$(k_s)$ and Im$(k_n)$ are found to be \emph{non-positive} at all frequencies, as required~\cite{Fatemeh2017} by the dissipative nature of lossy (e.g. frictional) contacts. 

\item As expected, application of normal stress to the fracture ($p\!=\!1.5$MPa) results in an overall increase of both normal and shear specific stiffness. This is accompanied by an apparent ``shortening'' of the fracture, also seen in the discontinuity maps in Fig.~\ref{geomtry_recons}.

\item The uncertainty of the recovered $k_\alpha$-profiles generally increases toward the (apparent) fracture tip, as supported by the vanishing FOD in that direction.

\item The imaginary parts of $k_s$ and $k_n$ suggest frictional energy dissipation at the interface, whose apparent increase with~$p$ could be attributed to the growing contact area. 

\item For the granite specimen tested, the use of nominal \emph{vs.} reconstructed modulus does not yield significant differences in terms of the obtained 
$k_s$ and $k_n$~profiles. 
\end{itemize}

\section{Conclusions}
 
\noindent The objective of this study is to establish a laboratory framework and data analysis platform for the full-field, quantitative ultrasonic characterization of heterogeneous fractures in rock. To this end, O(10$^4$Hz) waves are propagated in a slab-like rock specimen, using wavelengths sufficiently long to ensure the fidelity of the plane-stress approximation that provides a lynchpin for the analysis. Thus generated in-plane wavefield is monitored by a Scanning Laser Doppler Vibrometer (SLDV), which upon suitable signal processing furnishes spatially- and temporally-differentiable displacement waveforms over the prescribed scanning area. Such full-field sensory data are then used to (i) compute the maps of elastic modulus in the rock specimen (before and after fracturing) via a technique known as elastography; (ii) identify the fracture geometry as the support of persistent spatial discontinuities in the transient SLDV fields; (iii) expose the fracture's primal (traction-displacement jump) contact behavior, and (iv) identify via Fourier analysis its effective profiles of complex-valued, shear and normal specific stiffness. The results are verified for self-consistency at several key stages of the analysis, and are found to conform with expected trends in terms of the interfacial fracture response to seismic waves. The fracture sensing methodology is expected to work over a sizable range of excitation frequencies as it indiscriminately considers body and interfacial waves, and folds equally the amplitude and phase information into the analysis.  The proposed full-field approach, wich furnishes high-resolution recovery of the heterogeneous shear and normal specific stiffness in fractured laboratory specimens, opens the door toward: (a) better understanding of the intricate relationship between fracture's multi-scale geometry, aperture, interphase properties, and its mechanical (conservative as well as dissipative) characteristics, and (b) commensurate in-depth analysis of fracture networks under seismic excitation, where the complexity of underlying wave interaction makes such systems largely resilient to remote sensing. Another unique opportunity provided by the present framework is that of exposing the relationship between the advancing process zone (observed e.g. via acoustic emission) in strained laboratory specimens and its posterior ``damage contrail'', as sensed by ultrasonic waves. 

\section*{Acknowledgments} 

\noindent The corresponding author kindly acknowledges the support provided by the National Science Foundation (Grant CMMI-1536110) and the University of Minnesota's Supercomputing Institute. The data supporting the analysis and conclusions can be accessed by contacting the lead author.

\bibliography{inverse,crackbib}

\end{document}